\documentclass[reprint,prl,twocolumn,superscriptaddress,showpacs,showkeys,floatfix,preprintnumbers]{revtex4-2}
\usepackage[english]{babel}
\usepackage[utf8]{inputenc}
\usepackage{float}
\usepackage{amsmath,amssymb,graphicx,textcomp}
\usepackage{hyperref}
\usepackage[caption=false]{subfig}
\usepackage{tikz-feynman}
\tikzfeynmanset{compat=1.0.0}

\newcommand*\dif{\mathop{}\!\mathrm{d}}

\begin{document}

\date{July 30, 2022}



\title{Lattice QCD calculation of $\pi^0\rightarrow e^+ e^-$ decay}

\newcommand{\UCONN}{
  Physics Department,
  University of Connecticut,
  Storrs, Connecticut 06269-3046,
  USA}

\newcommand{\CU}{
  Physics Department,
  Columbia University,
  New York, New York 10027,
  USA}

\newcommand{\PKU}{
  School of Physics,
  Peking University,
  Beijing 100871, 
  China}
  
\newcommand{\CICQM}{
  Collaborative Innovation Center of Quantum Matter, 
  Beijing 100871, China}
   
\newcommand{\CHEP}{
Center for High Energy Physics, 
Peking University, 
Beijing 100871, China}

\author{Norman Christ}
\affiliation{\CU}

\author{Xu Feng}
\affiliation{\PKU, \CICQM, \CHEP}

\author{Luchang Jin}
\affiliation{\UCONN}

\author{Cheng Tu}
\affiliation{\UCONN}

\author{Yidi Zhao}
\affiliation{\CU}

\begin{abstract}

We extend the application of lattice QCD to the two-photon-mediated, order $\alpha^2$ rare decay $\pi^0\rightarrow e^+ e^-$.  By combining Minkowski- and Euclidean-space methods we are able to calculate the complex amplitude describing this decay directly from the underlying theories (QCD and QED) which predict this decay.  The leading connected and disconnected diagrams are considered; a continuum limit is evaluated and the systematic errors are estimated.  We find $\mathrm{Re} \mathcal{A} = 18.60(1.19)(1.04)\,$eV, $\mathrm{Im} \mathcal{A} = 32.59(1.50)(1.65)\,$eV, a more accurate value for the ratio $\frac{\mathrm{Re} \mathcal{A}}{\mathrm{Im} \mathcal{A}}=0.571(10)(4)$ and a result for the partial width $\Gamma(\pi^0\to\gamma\gamma) = 6.60(0.61)(0.67)\,$eV.  Here the first errors are statistical and the second systematic.  This calculation is the first step in determining the more challenging, two-photon-mediated decay amplitude that contributes to the rare decay $K\to\mu^+\mu^-$.
\end{abstract}

\maketitle

{\em Introduction}. --- Recent advances in the methods of lattice QCD allow the calculation of increasingly complex processes which involve both QCD and QED.  While initially motivated by the calculation of the light-by-light scattering process which contributes to the anomalous magnetic moment, $g_\mu-2$,  of the muon~\cite{Blum:2015gfa, Green:2015mva}, these methods can be applied to other processes where an {\it ab initio} lattice QCD result would be of value.  Of particular interest is the rare flavor-changing neutral current (FCNC) decay $K_L \rightarrow \mu^+ \mu^-$.  This decay results from both an accurately-calculated short-distance second-order weak process and a largely unknown electroweak process involving a two-photon intermediate state.  A  comparison of the second-order weak prediction with experiment is important for testing the standard model and exploring physics beyond it. However, to compare this standard model prediction with experiment, we require a first-principles calculation of the long-distance, two-photon contribution to this decay.  

In contrast with light-by-light scattering entering $g_\mu-2$ which can be computed using Euclidean-space methods, $K_L \rightarrow \mu^+ \mu^-$ is inherently a Minkowski process with a complex amplitude resulting from physical time evolution.  Here we develop a method that allows us to deal with such a process using lattice QCD and apply this method to calculate the complex $\pi^0 \rightarrow e^+ e^-$ decay amplitude --- a process which can be viewed as a simpler version of the $K_L \rightarrow \mu^+ \mu^-$ decay but suffering from the same problem of a potentially on-shell, two-photon intermediate state.  It is the lattice QCD calculation of the $\pi^0 \rightarrow e^+ e^-$ decay which is the subject of this paper.

\begin{figure}
\includegraphics[width=0.4\textwidth]{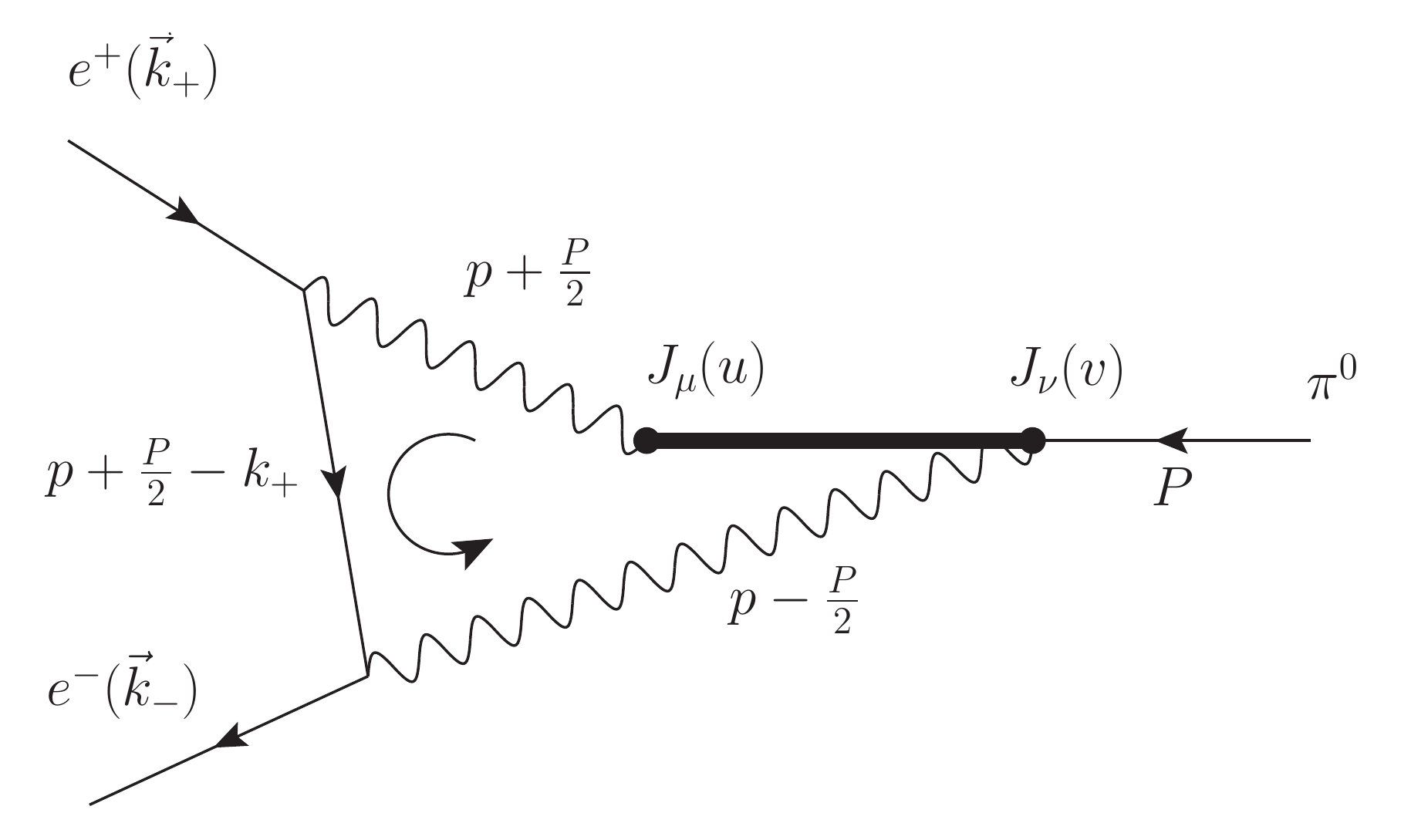}
\caption{The leading order Feynman diagram for $\pi^0 \rightarrow e^+ e^-$ decay.  Here we do not show the quark structure of the QCD portion of the amplitude which is represented by the heavy line.  However, the two hadronic electromagnetic currents are distinguished and their time ordering suggested.}
\label{fig:pi2ee}
\end{figure}

To leading-order the $\pi^0 \rightarrow e^+ e^-$ decay is mediated by a two-photon intermediate state as shown in Fig.~\ref{fig:pi2ee}. This decay is highly suppressed by the chiral symmetry of the electromagnetic interaction of the electron which results in a decay rate proportional to $(m_e/M_\pi)^2$ where $m_e$ and $M_\pi$ are the masses of the electron and $\pi^0$.   

This decay amplitude $\mathcal{A}$ is complex with an imaginary part determined by the optical theorem, giving the well-known unitary bound for the $\pi^0\rightarrow e^+ e^-$ branching ratio.  We will use lattice QCD to compute both the real and imaginary parts of $\mathcal{A}$ using a treatment of QED without power-law finite-volume errors.  Thus, our result for the imaginary part provides an improved calculation of the partial width $\pi^0\to\gamma\gamma$ with finite volume errors that vanish exponentially with increasing volume, without the constraint to use photon momenta that obey quantization conditions related to the volume used in the lattice QCD calculation~\cite{Meyer:2013dxa}.  (Preliminary results from these methods appear in Refs.~\cite{Christ:2020bzb, Christ:2020dae, Meng:2021ecs, Meng:2021las}.)

The $\pi^0 \rightarrow e^+ e^-$ decay has been measured to near 1\% accuracy in the KTeV experiment~\cite{Abouzaid:2006kk}.  After the contribution of $e^+ e^-\gamma$ Dalitz decays have been removed, the conventional $O(\alpha)$ radiative corrections must be performed. These $O(\alpha)$ corrections~\cite {Vasko:2011pi, Husek:2014tna} require a two-loop calculation in which the zeroth-order $\pi^0\to e^+ e^-$ decay is not approximated as point-like.  As is conventional, we use the calculation of Refs.~\cite {Vasko:2011pi, Husek:2014tna} to remove these radiative effects from the experimental results, giving $B\left( \pi^0\rightarrow e^+ e^- \right)  = 6.86(27)_\mathrm{stat.}(23)_\mathrm{syst.} \times 10^{-8}$.  This modified experimental result can be compared with our lattice QCD calculation which also does not include these corrections.  These corrections depend on a low-energy constant $\chi^{(r)}=4.5$, determined self-consistently from the experimental result with these corrections removed.  

This corrected experimental value is larger than most theoretical results including that obtained here, $B\left( \pi^0\rightarrow e^+ e^- \right) = 6.22(5)_\mathrm{stat}(2)_\mathrm{syst}\times 10^{-8}$.  See Ref.~\cite{Hoferichter:2021lct} for a recent precise theoretical prediction and extensive references to earlier results.

{\em Computational strategy}. --- We start with the Minkowski-space expression for the decay amplitude:
\begin{widetext}
\begin{eqnarray}
\mathcal{A} &=&e^4\int \dif^4w \, \langle 0|T\bigl\{J_\mu\bigl(\frac{w}{2}\bigr)J_\nu\bigl(-\frac{w}{2}\bigr)\bigr\}|\pi^0\rangle
\hskip 0.5 in \mbox{\ }  \label{eq:pi-MS} \\
&&\int \frac{\dif^4p}{(2\pi)^4} e^{-i p \cdot w} 
                         \left[ \frac{g_{\mu\mu'}}{( p+\frac{P}{2})^2-i\epsilon}\right]
                         \left[ \frac{g_{\nu\nu'}}{( p-\frac{P}{2})^2-i\epsilon}\right] \nonumber \overline{u}(k_-,h)\gamma_{\mu'}\left[ \frac{\gamma\cdot(p+\frac{P}{2}-k_+) + m_e}{(p+\frac{P}{2}-k_+)^2+m_e^2-i\epsilon}\right]\gamma_{\nu'} v(k_+,h), \nonumber
\end{eqnarray}
\end{widetext}
where $w = u - v$ is the relative space-time position of the two electromagnetic currents, $P = (\vec{0}, M_\pi)$ is the four-momentum of initial pion and $h=\pm\frac{1}{2}$ is the helicity of the electron on which $\mathcal{A}$ does not depend. Note that we have integrated out their average position $\frac{u+v}{2}$ and removed the resulting delta function that imposes total energy and momentum conservation.  The Minkowski metric tensor $g_{\mu\mu'} =  \text{diag}(-1,1,1,1)$.

\begin{figure}[h]
\centering
\includegraphics[width=0.39\textwidth]{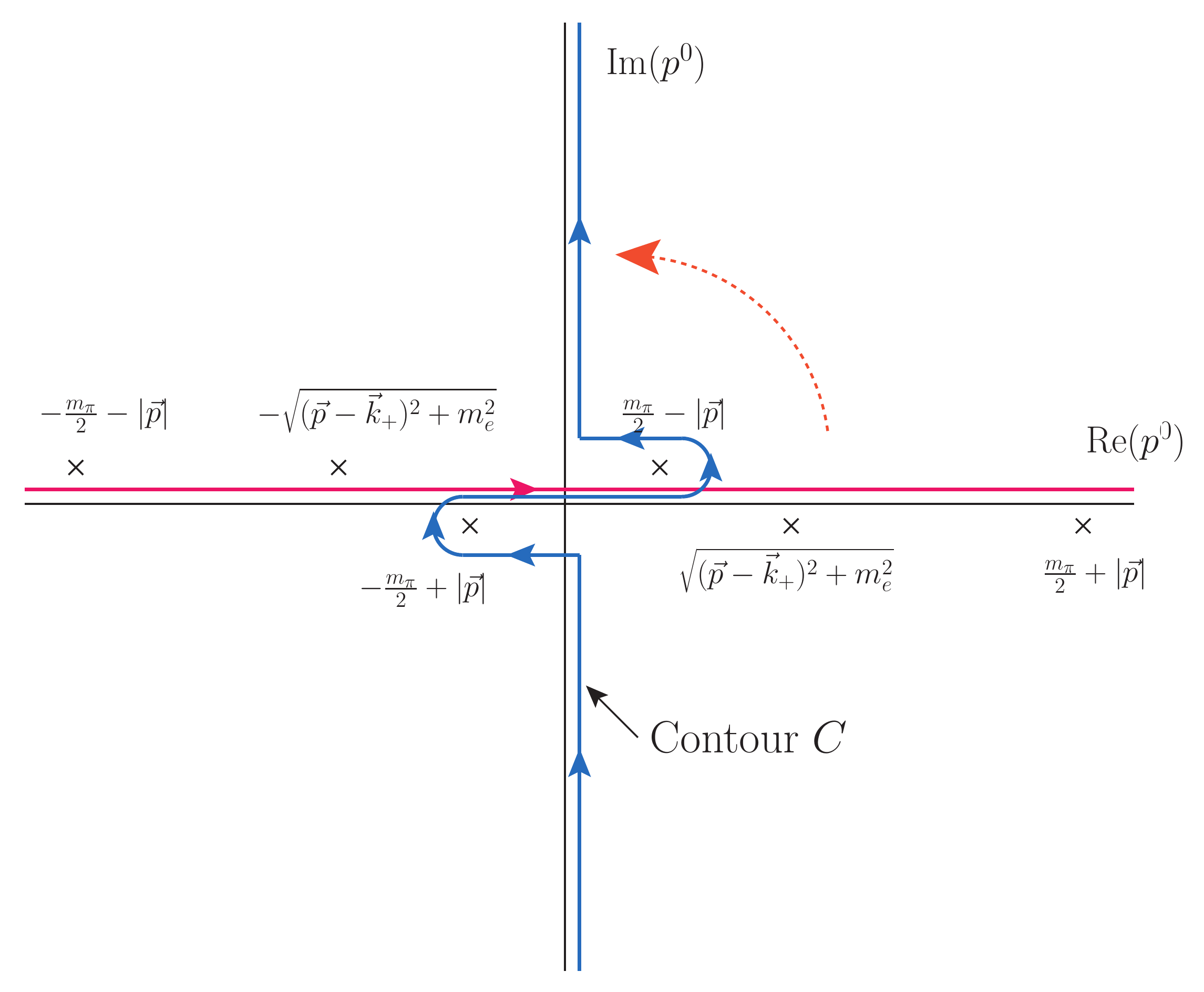}
\caption{The deformed $p^0$ integration contour. The six crosses locate the six poles of the integrand in Eq.~\eqref{eq:pi-MS}.}
\label{fig:p0-contour}
\end{figure}

A direct Euclidean-space calculation of the hadronic matrix element in Eq.~\ref{eq:pi-MS} using lattice methods is feasible if we Wick-rotate the $w^0$ contour by replacing the real variable $w^0$ with the product $e^{-i\phi}w_0$ and increasing $\phi$ from 0 to $\pi/2$ while keeping $w_0$ real, thereby making $w^0$ a Euclidean time. At the same time, the $p^0$ contour must also be rotated so that the exponent $i p^0 w^0$ in Eq.~\ref{eq:pi-MS} remains purely imaginary as $|p^0|\to\infty$. Because of the existence of the two-photon intermediate state whose energy may be lower than the energy of initial pion state, the rotated $p^0$ contour cannot simply follow the imaginary axis.  Instead the $p^0$ contour must be distorted as in Fig.~\ref{fig:p0-contour} to avoid the poles which cross the imaginary axis when the energy of the two-photon intermediate state is lower than the pion mass, $2 |\vec{p}| < M_\pi$.

This choice of contour guarantees that the $p^0$ integral will converge.  However, once the variable $w^0$ has become imaginary, the factor $e^{-i p \cdot w}$ will grow exponentially with $|w^0|$ for the real values of $p^0$ that appear for the contour of Fig.~\ref{fig:p0-contour}.  Fortunately, the Euclidean-space hadronic matrix element $H_{\mu\nu}(w)$ will compensate for this behavior, so that both the $p^0$ and $w^0$ integrations are convergent.  The dependence of the hadronic matrix element on $w^0$ can be determined by inserting a complete set of intermediate states between the two electromagnetic currents in the hadronic matrix element $\langle 0| T\bigl\{J_\mu(\frac{w}{2})J_\nu(-\frac{w}{2})\bigr\}|\pi^0\rangle_E$:
\begin{equation}
  \sum_n \langle 0 | J_\mu(\frac{\vec{w}}{2}, 0)|n\rangle\,\langle n|J_\nu(-\frac{\vec{w}}{2}, 0) |\pi^0\rangle e^{|w^0|{(\frac{M_\pi}{2}-E_n)}}.
\end{equation}
The lightest intermediate state is the two-pion state with $E_n = 2 M_\pi$. Thus, this Euclidean-space hadronic matrix element will decay as $\exp(-3M_\pi |w^0| / 2)$ for large $|w^0|$.  This large-$|w^0|$  fall-off is sufficient to overcome the $\exp{(|w^0|M_\pi/2)}$ growth due to the $p^0$ contour in Fig.~\ref{fig:p0-contour}.

We can express the analytic expression for decay amplitude after contour deformation as follows:
\begin{eqnarray}
  \mathcal{A} &=& \int \dif^4w \  L_{\mu\nu}(w) H_{\mu\nu}(w) \label{eq:factor} \\
L_{\mu\nu}(w) &=& e^4\int \frac{\dif^3p}{(2\pi)^4} \int_C \dif p^0\  e^{-i \vec p \cdot \vec w}  e^{+p^0 w^0} \nonumber \\
       &&\hskip -0.3 in \left[ \frac{\widetilde{g}_{\mu\mu'}}{( p+\frac{P}{2})^2-i\epsilon}\right]
       \left[ \frac{\widetilde{g}_{\nu\nu'}}{( p-\frac{P}{2})^2-i\epsilon}\right] \overline{u}(k_-,\frac{1}{2})\gamma_{\mu'} \nonumber \\
&&\hskip -0.1 in  \left[ \frac{\gamma\cdot(p+\frac{P}{2}-k_+) + m_e}{(p+\frac{P}{2}-k_+)^2+m_e^2-i\epsilon}\right]\gamma_{\nu'} v(k_+,\frac{1}{2}) \label{eq:L-factor} \\
H_{\mu\nu}(w) &=& 
   \langle 0|T\bigl\{J_\mu\bigl(\frac{w}{2}\bigr)J_\nu\bigl(-\frac{w}{2}\bigr)\bigr\}|\pi^0\rangle_E.
\label{eq:H-factor}
\end{eqnarray}
where $\widetilde{g}_{\mu\mu'} = \text{diag}(i,1,1,1)$ is introduced to connect the Minkowski conventions for the electromagnetic currents in $L_{\mu\nu}$ with the Euclidean conventions used in the hadronic matrix element. Note that the amplitude $\mathcal{A}$ is not altered by these changes of contour and remains complex with real and imaginary parts which can be computed from the Euclidean-space amplitude in Eq.~\eqref{eq:H-factor}.

The leptonic factor $L_{\mu\nu}$ is evaluated by performing the $p^0$ integral using Cauchy's theorem, resulting in a three-dimensional integral over $\vec{p}$. When integrated over $|\vec p\,|$ the singular factor $\frac{1}{|\vec{p}| - M_\pi / 2-i\epsilon}$ gives both real and imaginary parts.  Recognizing that the integral is independent of the direction of the outgoing positron momentum $\vec{k}_+$, allows us average over this direction.  Here we present the result for the spatial components, writing $L_{ij}(w) =  L(w^0,|\vec w|)\epsilon_{ijk} w^k/|\vec w|^2$ with
\begin{widetext}
\begin{align}
  L^{\text{im}}(w^0, |\vec w|) &=  2\pi \alpha^2\frac{m_e}{M_\pi^2 } \frac{1}{\beta} \ln \left(\frac{1 + \beta}{1 - \beta} \right) F(\frac{M_\pi}{2}|\vec{w}|) \label{eq:L_lep} \\
  L^{\text{re}}(w^0,|\vec w|) &= 4  m_e \alpha^2  \left\{\ln \left(\frac{1 + \beta}{1 - \beta}\right) \mathcal{P}\!\!\!\!\!\!\int_0^\infty \frac{\dif |\vec{p}|}{M_\pi^2\beta} \frac{e^{-|\vec{p}||w^0|}}{\left(\vec{p}\,^2-\frac{M_\pi}{2}\right)^2} F(|\vec{p}||\vec{w}|)\left[\frac{M_\pi}{2} \sinh(\frac{M_\pi}{2} |w^0|) + |\vec{p}|\cosh(\frac{M_\pi}{2} |w^0|)\right] \right.  \\
            & \hskip 1.6 in + \left.  \int_0^\infty\!\! \dif |\vec{p}| \dif \cos\theta \frac{e^{-E_e(|\vec{k}_+|,|\vec p|,\theta)|w^0|}F(|\vec{p}||\vec{w}|)}{E_e(|\vec{k}_+|,|\vec p|,\theta) \;          
            \bigl(-M_\pi + 2|\vec{k}_+|\cos\theta\bigr)\bigl(M_\pi+2|\vec{k}_+|\cos\theta\bigr)}  \right\}  \nonumber
.\end{align}
\end{widetext}
Here $\beta=\sqrt{1-4m_e^2/M_\pi^2}$, $F(x) = \cos(x)-\sin(x)/x$ and $E_e(k,p,\theta) = \sqrt{k^2+p^2-2pk\cos(\theta)+m_e^2}$.  We then evaluate the leptonic factor $L_{\mu\nu}$ as a two-dimensional numerical integral, requiring that the integration error lies below $0.001\%$. The result is tabulated as a function of $w^0$ and $|\vec{w}|$.  We evaluate the four-dimensional integral over $w$ in Eq.~\eqref{eq:factor} as a sum over lattice points with the values of $L_{\mu\nu}(w)$ obtained from this table by linear interpolation.

{\em Hadronic matrix element}. ---The hadronic matrix element $H_{\mu\nu}$ in Eq.~\eqref{eq:H-factor} can be calculated from the three-point function:
\begin{eqnarray}
  \langle 0 | T \bigl\{J_\mu(x) J_\nu(0) \bigr\}| \pi \rangle && \label{eq:3pt} \\
&&  \hskip -1.0 in =  \frac{Z_V^2}{N_\pi} \underset{t\rightarrow -\infty}{\lim}  e^{M_\pi |t|} \bigl\langle J_\mu(x) J_\nu(0) \pi^0(t) \bigr\rangle^\mathrm{lat}, 
\nonumber
\end{eqnarray}
which can be computed using lattice QCD.  It is only in this evaluation of  $H_{\mu\nu}$ that a finite volume is introduced.  Since this Euclidean-space amplitude involves no massless particles, all finite-volume errors will decrease exponentially in the linear size of the volume.

The factor $Z_V$ renormalizes the non-conserved, local current used in the lattice calculation of the right hand side of Eq.~\eqref{eq:3pt}.  The factor $N_\pi$ is obtain from the $\pi^0$ two-point function and compensates for the normalization of the pion interpolating operator $\pi^0(t)$. We use a Coulomb-gauge-fixed wall source for this operator.

\begin{figure}
\hskip 0.18 in \subfloat[Connected]{
\centering
\includegraphics[width=0.18\textwidth]{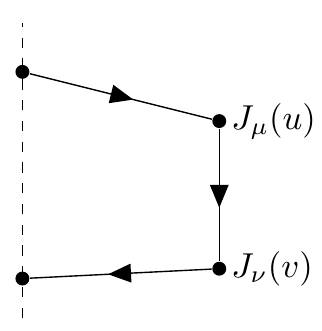}
  \label{fig:connected}
}
\hskip 0.18 in
\subfloat[Disconnected]{
\centering
\includegraphics[width=0.18\textwidth]{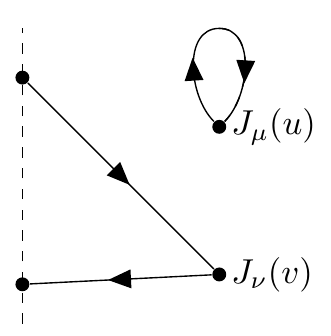}
    \label{fig:disconnected}
}
\caption{The quark propagator contractions needed for the hadronic three-point function given in Eq.~\eqref{eq:3pt}. The dashed line locates the wall-source pion interpolating operator.} \label{fig:feyn}
\end{figure}

The two diagrams needed to evaluate this three-point function are shown in Fig.~\ref{fig:feyn}. The first is connected and can be constructed from two wall-source propagators and one point-source propagator. The two electromagnetic currents must be located sufficiently far from the pion interpolating operator that contamination from states more energetic than the pion can be neglected. We require the pion source at $t$ to be separated from the closer current by a fixed time $\Delta t$.  Let $T$ be time extent of the lattice and require that all time coordinates lie in the interval $[0,T)$.  Label the time locations of current operators $t_>$ and $t_<$ where $(t_>-t_<)\%T < T/2$.   We then average the results from two choices for the time $t$ of the pion wall source.  For the first we chose $t=(t_<-\Delta t)\%T$ and for the second $t=(t_>+\Delta t)\%T$.

The second diagram is quark-line-disconnected. Such disconnected diagrams typically involve large statistical noise and are difficult to calculate. In our calculation, we use for the quark loop shown in Fig.~\ref{fig:disconnected} the results for $\mathrm{Tr}\left[D^{-1}(x, x)\gamma_\mu\right]$ computed using all-to-all propagators computed with randomly-displaced, $3^4$ grid sources from the RBC/UKQCD calculation of the disconnected contribution to the hadronic vacuum polarization component of $g_\mu-2$~\cite{Blum:2015you}. As shown in Tab.~\ref{tab:24ID_disc}, we determine this disconnected amplitude with a statistical error of 60\%.

The hadronic matrix element is calculated on four ensembles, whose parameters are listed in Tab.~\ref{tab:ensembles}. All ensembles use M\"{o}bius domain wall fermions (DWF), which achieve good chiral symmetry with a much smaller size in the 5th dimension than required by the conventional DWF action. All ensembles use the Iwasaki gauge action.  The 24ID, 32ID and 32IDF ensembles use the dislocation suppressing determinant ratio action to reduce chiral symmetry breaking effects. For every configuration, we have 1024 or 2048 point-source propagators with randomly distributed sources and Coulomb-gauge-fixed wall-source propagators with sources on every time slice. 

\begin{table}[thp]
  \small
  \centering
  \begin{tabular}{|c|c|c|c|c|c|c|}
    \hline
& 24ID & 32ID & 32IDF & 48I & 64I \\
\hline
$a^{-1}$ (GeV) & 1.015 & 1.015 & 1.37& 1.73 & 2.36 \\
$M_\pi$ (MeV) & 140 & 140  & 143 & 135 & 135\\
Configuration separation & 10 & 10  & 10 & 10 & 20\\
Configurations & 47 & 47  & 61 & 32 & 49\\
point sources & 1024 & 2048 & 1024 & 1024 &1024 \\
$\Delta t$ & 10 & 10 & 14 & 16 & 22\\
\hline
  \end{tabular}
  \caption{Table of gauge-field ensembles used here. These were generated by the RBC/UKQCD collaborations~\protect\cite{Blum:2014tka}.}
  \label{tab:ensembles}
\end{table}

\begin{figure}
\subfloat[Imaginary part]{
    \includegraphics[width=0.43\textwidth]{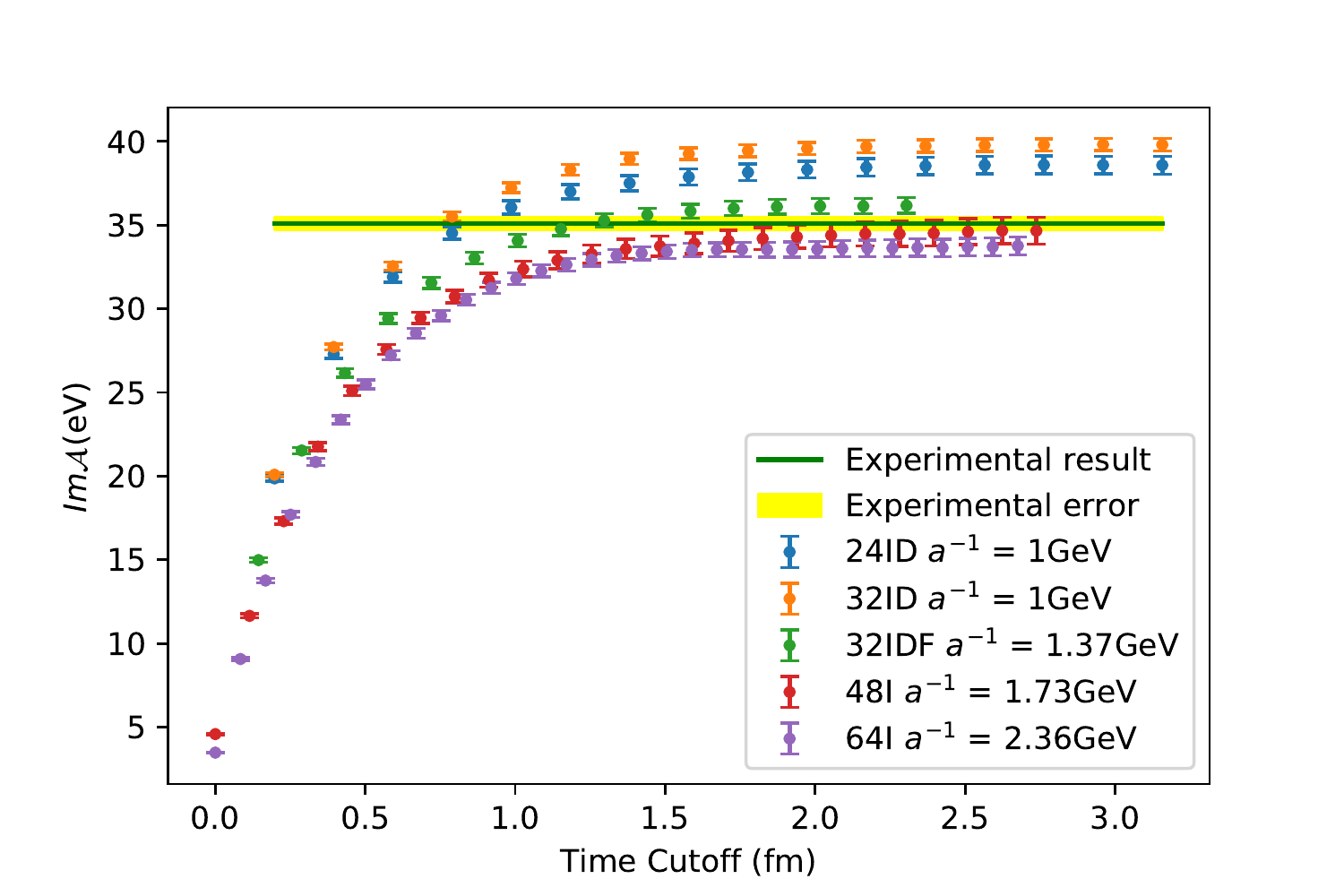}
    \label{fig:imag}
}
  \hfill
\subfloat[Real part]{
    \includegraphics[width=0.43\textwidth]{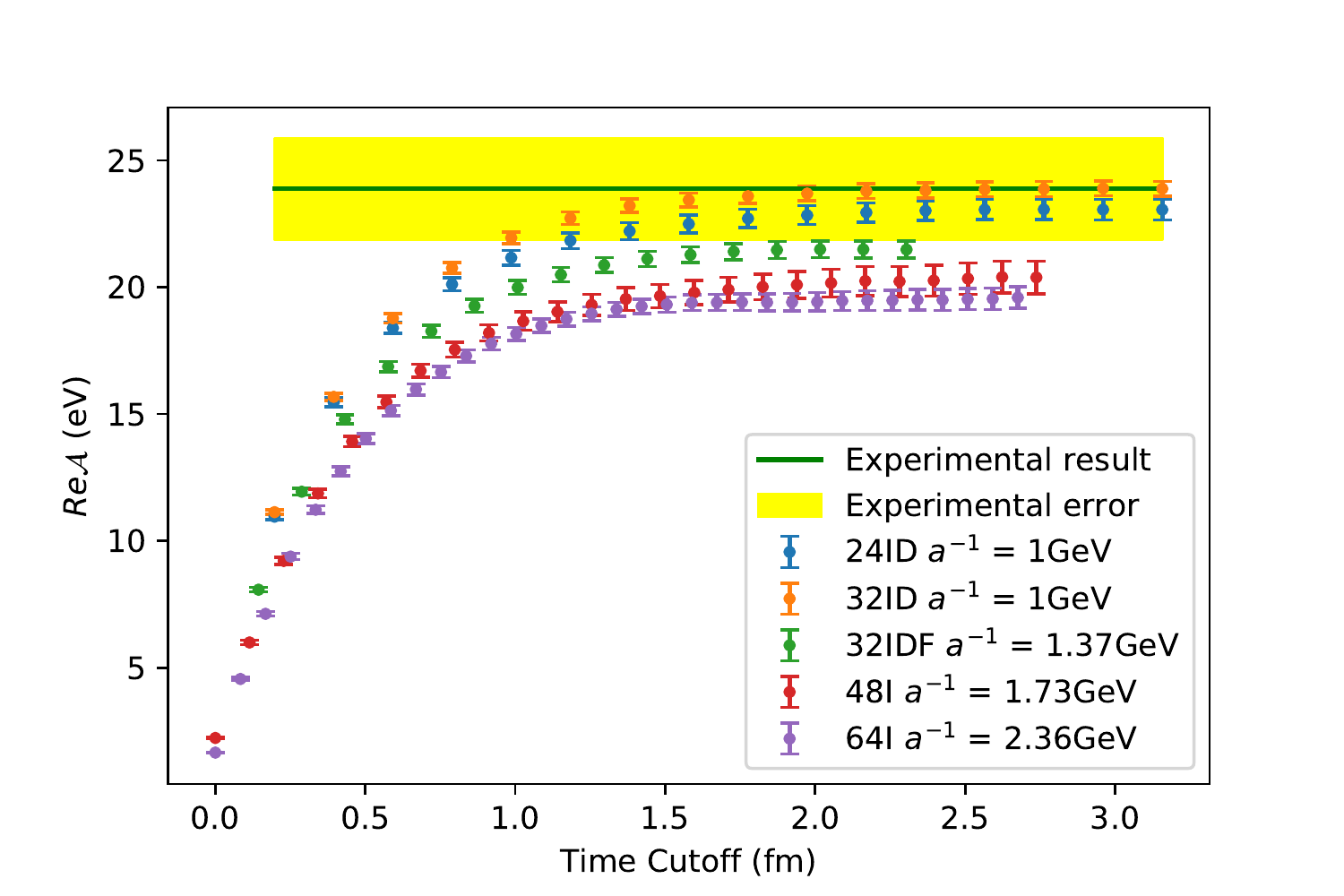}
   \label{fig:real}
}
  \caption{The decay amplitude plotted as a function of $t$. We sum over the difference $x^0$ between the two times at which the currents in Eq.~\eqref{eq:3pt} are evaluated, requiring $|x^0|\le t$.}
  \label{fig:amplitude}
\end{figure}

{\em Results}. ---The calculated real and imaginary parts of the amplitude are listed in Tab.~\ref{tab:amplitudes} and plotted in Fig.~\ref{fig:amplitude}. In Tab.~\ref{tab:amplitudes} the experimental value for the imaginary part is evaluated using the optical theorem and the experimental pion life time; the experimental real part is obtained by subtracting the imaginary part contribution from the experimental decay rate, with radiative corrections. In the table and plot only the contribution from the connected diagram is included.  The contribution from the disconnected diagram is treated as a source of systematic error. The disconnected diagram is calculated for the 24ID ensemble and the result shown in Tab.~\ref{tab:24ID_disc}.

\begin{table}[thp]
  \small
  \begin{center}
    \begin{tabular}{|c|c|c|c|c|c|}
      \hline
Source & Im $\mathcal{A}$ (eV) & Re $\mathcal{A}$ (eV) \\ \hline
      24ID &38.58(54) & 23.06(40)  \\
      32ID & 39.80(36) & 23.88(29)  \\
      32IDF & 36.17(47) & 21.48(33)  \\
      48I & 34.66(80) & 20.38(65) \\ 
      64I & 33.75(53) & 19.59(42) \\ \hline
      Experiment &35.07(37) & 24.1(2.0) \\
      \hline
    \end{tabular}
  \end{center}
  \caption{The lattice and experimental results for the real and imaginary parts of the decay amplitude in eV. The error in parenthesis is statistical or experimental. }
  \label{tab:amplitudes}
\end{table}

\begin{table}[thp]
  \small
  \begin{center}
    \begin{tabular}{|c|c|c|c|c|c|}
      \hline
Diagram & $\text{Im}\mathcal{A}$ (eV) & $\text{Re}\mathcal{A}$ (eV) \\ \hline
      Connected &38.58(54) & -23.06(40)  \\
      Disconnected & -1.11(55) & 0.62(40)  \\
      \hline
    \end{tabular}
  \end{center}
  \caption{The contribution to amplitude from connected and disconnected diagrams for the 24ID ensemble. The errors in parenthesis are statistical.}
  \label{tab:24ID_disc}
\end{table}

\begin{figure}[t]
\subfloat[Imaginary part]{
    \includegraphics[width=0.43\textwidth]{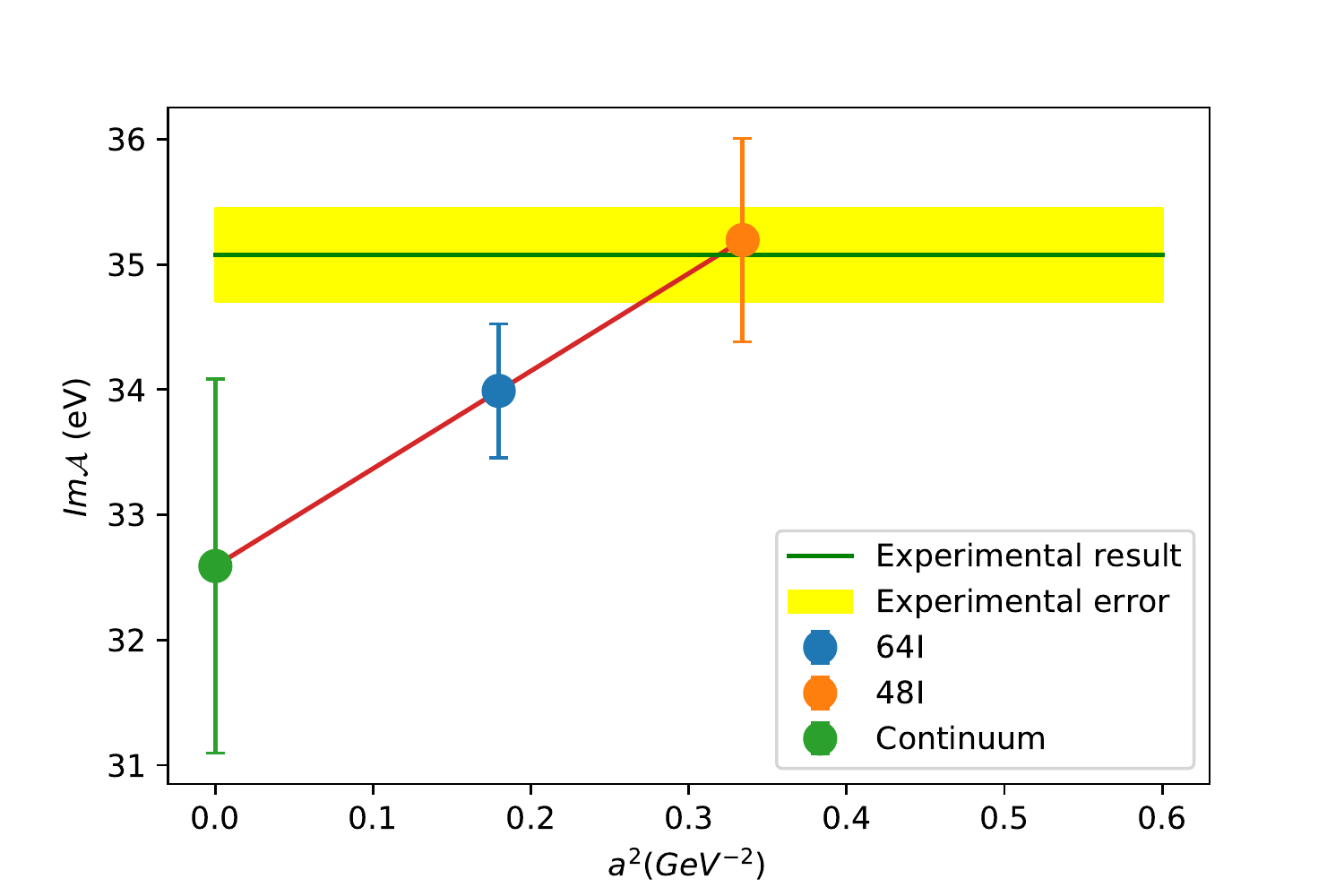}
    \label{fig:imag}
}
  \hfill
\subfloat[Real part]{
    \includegraphics[width=0.43\textwidth]{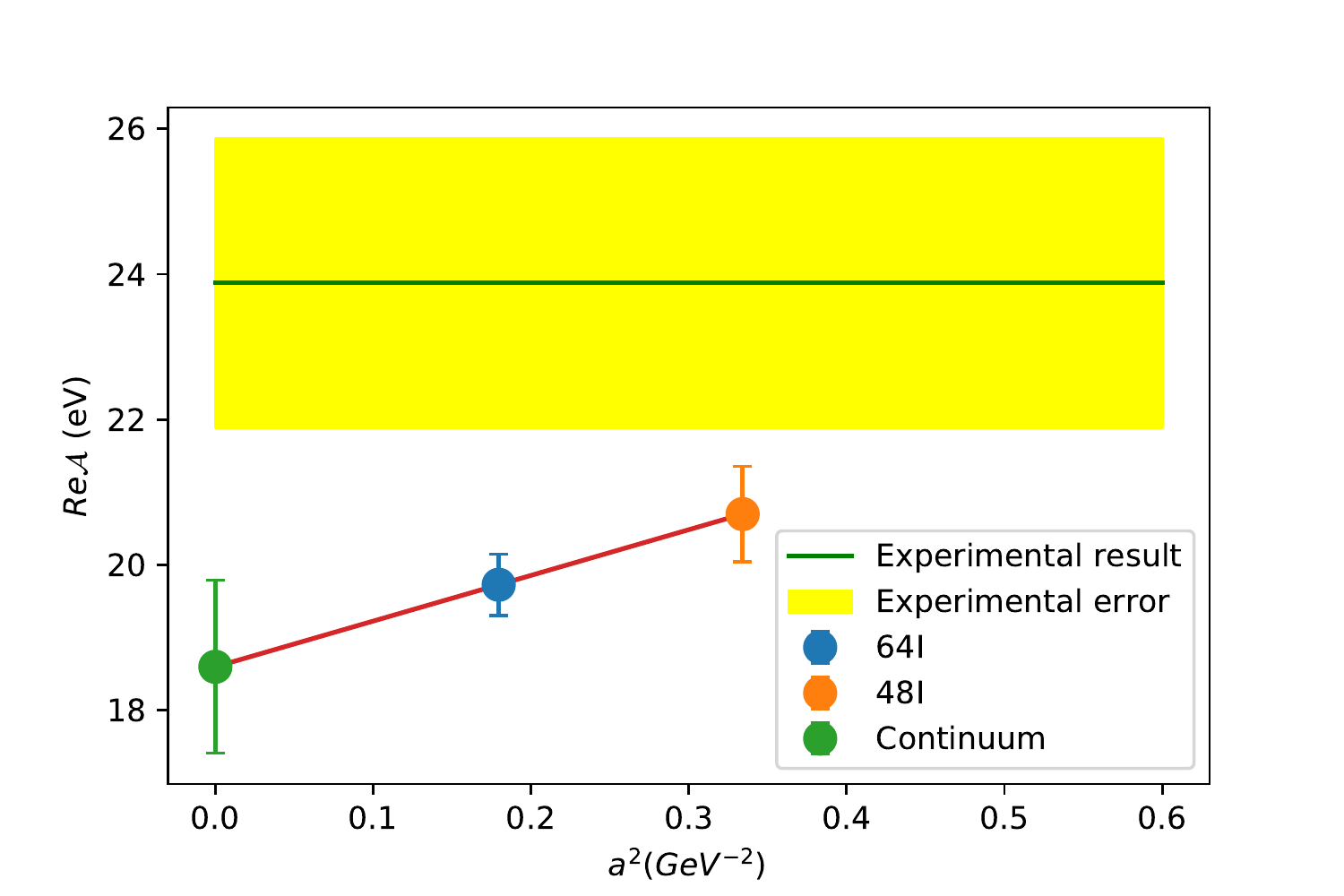}
    \label{fig:real}
}
  \caption{Extrapolation to the continuum limit showing statistical errors.}
  \label{fig:extrapolation}
\end{figure}

We use the continuum limit extrapolated from the 48I and the 64I ensembles as shown in Fig~\ref{fig:extrapolation} as our final result.  Estimates of the dominant systematic errors are presented in Tab.\ref{tab:sys-error}. The finite-volume error is estimated by evaluating the difference between the 24ID and 32ID results and assuming that it behaves as $e^{-M_\pi L}$, where $L$ is the linear spatial lattice size. The error from omitting disconnected diagram is estimated by comparing the contributions of connected and disconnected diagrams for the 24ID ensemble listed in Tab.~\ref{tab:24ID_disc}.   The error resulting from our pion mass being 4 MeV larger than physical is estimated from chiral perturbation theory while that arising from our choice of $Z_V$ is taken from Ref.~\cite{Blum:2014tka}.

\begin{table}[htp]
  \small
  \begin{center}
    \begin{tabular}{|c|c|c|c|c|c|}
      \hline
Sources & Im $\mathcal{A}$ (eV) & Re $\mathcal{A}$ (eV) \\ \hline
Finite volume &  1.21 & 0.81  \\
Disconnected diagram & 1.00 & 0.55 \\
Unphysical pion mass  &  0.3  & 0.3  \\
$Z_V$ & 0.024 & 0.014 \\ \hline
Total systematic error& 1.57 & 0.98 \\
      \hline
    \end{tabular}
  \end{center}
  \caption{Estimated systematic errors in percent.}
  \label{tab:sys-error}
\end{table}

Our final results are Im$\mathcal{A} = 32.59(1.50)_{\mathrm{stat}}(1.65)_{\mathrm{sys}}$ eV, Re$\mathcal{A} = 18.60(1.19)_{\mathrm{stat}}(1.04)_{\mathrm{sys}}$ eV and
\begin{equation}
\frac{\text{Re} \mathcal{A}}{\text{Im} \mathcal{A}} = 0.571(10)_\mathrm{stat}(4)_{\mathrm{sys}}.
\end{equation}
The smaller error on this ratio results from statistical correlations and our method of estimating the systematic errors.   We can combine our more accurate result for this ratio with the experimental result for the decay width $\Gamma(\pi^0\to\gamma\gamma)$ to obtain more accurate values for the real part of the decay amplitude and the branching ratio\footnote{We thank both P. Sanchez and P. Masjuan and also  A.~Soni for independently suggesting this error-reduction strategy}:
\begin{eqnarray}
\mathrm{Re}\mathcal{A} =  20.2(0.4)_\mathrm{stat}(0.1)_\mathrm{syst}(0.2)_\mathrm{expt}\,\mathrm{eV} \\
B(\pi^0\to e^+e^-) =  6.22(5)_\mathrm{stat}(2)_\mathrm{syst}\times 10^{-8}.
\end{eqnarray}
The error labeled ``expt'' arises from the error on the measured $\pi^0\to\gamma\gamma$ decay rate.  This result for the branching ratio is 1.8 sigma below the experimental value~\cite{Abouzaid:2006kk} $B(\pi^0\to e^+e^-)_{\mathrm{expt}} = 6.86(27)_\mathrm{stat.}(23)_\mathrm{syst.} \times 10^{-8}$, from which radiative corrections have been removed.   

Finally we can use our result for the imaginary part of $\mathcal{A}$ to determine a lattice QCD prediction for the decay width $\Gamma(\pi^0\to\gamma\gamma)=6.60(0.61)_\mathrm{stat}(0.67)_\mathrm{syst}\,$eV to be compared with the experimental result~\cite{PrimEx-II:2020jwd} $\Gamma(\pi^0\to\gamma\gamma)=7.802(0.052)_\mathrm{stat}(0.105)_\mathrm{syst}\,$eV.  This new lattice QCD result is computed with physical quark masses, contains finite volume errors which are suppressed exponentially in the linear extent of the lattice and can be viewed as a refinement of earlier lattice results~\cite{Feng:2012ck, Gerardin:2019vio}.

{\em Conclusion and Outlook}. --- We have applied a combination of covariant Feynman perturbation theory and lattice QCD to calculate the complex amplitude describing the decay $\pi^0\to e^+e^-$.  Our result for this decay branching ratio is accurate at the 1\% level and follows the pattern of previous theoretical results lying below the experimental value.  Our method for combining lattice QCD with photon and lepton propagators extends techniques developed to calculate the hadronic light-by-light scattering contribution to $g_\mu-2$ and holds promise for the eventual calculation of the two-photon-exchange contribution to the rare decay $K_L\to\mu^+\mu^-$.

{\em Acknowledgments}. --- We would like to thank P. Sanchez and P. Masjuan as well as A. Soni for a valuable suggestion and our other RBC and UKQCD collaborators for helpful discussions and support. The calculations reported here were carried out on facilities of the USQCD Collaboration  funded by the Office of Science of the U.S. Department of Energy.  These calculations used gauge configurations and propagators created using resources of the Argonne Leadership Computing Facility, which is a DOE Office of Science User Facility supported under Contract DE-AC02-06CH11357. NHC and YZ were supported in part by US DOE grant \#DE-SC0011941.  XF was supported in part by NSFC of China under Grants No. 12125501, No. 12070131001, and No. 12141501, and National Key Research and Development Program of China under
No. 2020YFA0406400. L.C.J. acknowledges support by DOE Office of Science Early Career Award No. DE-SC0021147 and DOE Award No. DE-SC0010339.

\end{document}